# Observation of Spin Splitting in Room-Temperature Metallic Antiferromagnet CrSb

Meng Zeng[1]*, Ming-Yuan Zhu[1]*, Yu-Peng Zhu[1], Xiang-Rui Liu[1], Xiao-Ming Ma[1], Yu-Jie Hao[1], Pengfei Liu[1], Gexing Qu[2], Yichen Yang[3], Zhicheng Jiang[4], Kohei Yamagami[5], Masashi Arita[6], Xiaoqian Zhang[8], Tian-Hao Shao[1], Yue Dai[1], Kenya Shimada[6], Zhengtai Liu[3], Mao Ye[3], Yaobo Huang[7], Qihang Liu[1], Chang Liu[1]†

[1]Department of Physics and Shenzhen Institute for Quantum Science and Engineering (SIQSE), Southern University of Science and Technology (SUSTech), Shenzhen, Guangdong 518055, China
[2]Beijing National Laboratory for Condensed Matter Physics and Institute of Physics, Chinese Academy of Sciences, Beijing 100190, China
[3]State Key Laboratory of Functional Materials for Informatics, Shanghai Institute of Microsystem and Information Technology, Chinese Academy of Sciences, Shanghai 200050, China
[4]National Synchrotron Radiation Laboratory and School of Nuclear Science and Technology, University of Science and Technology of China, Hefei, Anhui 230026, China
[5] Japan Synchrotron Radiation Research Institute (JASRI), Sayo, Hyogo 679-5198, Japan
[6]Hiroshima Synchrotron Radiation Centre, Hiroshima University, Higashi-Hiroshima, Hiroshima 739-0046, Japan
[7]Shanghai Synchrotron Radiation Facility, Shanghai Advanced Research Institute, Chinese Academy of Sciences, Shanghai 201204, China
[8]Key Laboratory of Quantum Materials and Devices of Ministry of Education, School of Physics, Southeast University, Nanjing Jiangsu 211189, China

*These authors contribute equally to this work.
†Corresponding author. E-mail: liuc@sustech.edu.cn

# 1. Abstract

Recently, unconventional antiferromagnets that enable the splitting of electronic spins have been theoretically proposed and experimentally realized, where the magnetic sublattices containing moments pointing at different directions are connected by a novel set of symmetries. Such spin splitting (SS) is substantial, *k*-dependent, and independent of the spin-orbit coupling strength, making these magnets promising materials for antiferromagnetic spintronics. Here, combined with angle-resolved photoemission spectroscopy (ARPES) and density functional theory (DFT) calculations, we perform a systematic study on CrSb, a metallic spin-split antiferromagnet candidate with $T_N$ = 703 K. Our data reveals the electronic structure of CrSb along both out-of-plane and in-plane momentum directions, which renders anisotropic *k*-dependent SS and agrees well with the calculational results. The magnitude of such SS reaches up to at least 0.8 eV at non-high-symmetry momentum points, which is significantly higher than the largest known SOC-induced SS. This compound expands the choice of materials in the field of antiferromagnetic spintronics and is likely to stimulate subsequent investigations of high-efficiency spintronic devices that are functional at room temperature.

# 2. Introduction

In the framework of group theory, nonmagnetic materials without spin-orbit coupling (SOC), nonmagnetic materials with SOC, and magnetic materials with SOC can be fully characterized by space groups, double space groups, and magnetic double space groups, respectively. However, the symmetry of magnetic materials with negligible SOC is rarely explored. These materials possess a crucial property that the spin degree of freedom is partially decoupled from the orbital part, resulting in a substantial, SOC-independent, and momentum-dependent SS, the texture of which remains scarcely measured experimentally. To establish the theoretical underpinnings of such innovative antiferromagnets, Wu *et al.* and Hayami *et al.* proposed that the anisotropic kinetic motion of electrons gives rise to an effective SOC, resulting in an anisotropic SS [1–3]. Yuan *et al.* identified that significant *k*-dependent SS can exist in centrosymmetric, low-Z antiferromagnets[4]. They categorized the 1651 magnetic space groups into seven prototypes (SST-1 to SST-7) and predicted the presence of abundant AFM-induced SS materials within SST-4[5,6]. Several research groups introduced the concept of the spin space group[7–15] to describe the electronic structure of such materials and predicted a series of associated physical phenomena, such as the weak SOC $Z_2$ topological phase[7], the chiral Dirac-like fermions[16,17], the *C*-pair spin valley locking[18], the non-relativistic spin Hall effect[19,20], the spin splitter torque[21,22], the non-relativistic Edelstein effect[23], and the anomalous Hall effect[24,25]. Specifically, the term "altermagnet" is introduced to designate *collinear* antiferromagnets that possess such AFM-induced SS[26-28].

In recent years, unconventional antiferromagnetism has gained significant traction and emerged as a promising research field. Fruitful findings have been unveiled, such as the anomalous Hall effects in $RuO_2$[24] and MnTe[29], and the plaid-like spin splitting in $MnTe_2$[30]. However, certain constraints exist within the materials mentioned above. For example, despite substantial progress in understanding the unconventional antiferromagnetic nature of $RuO_2$[21,22,24,31], there is still controversy about its true magnetic phase [32–34]. Additionally, MnTe and $MnTe_2$, both of which are semiconductors with band gaps[30,35,36], exhibit suboptimal conductivity compared to metals, limiting their potential applications in spintronics.

In this article, we focus on CrSb, a candidate of unconventional antiferromagnet with Néel temperature significantly surpassing the room temperature ($T_N$ = 703 K for the bulk[37]). Compared to MnTe, CrSb is an out-of-plane *A*-type AFM metal, offering better conductivity and magnetic storage density. Compared to RuO$_2$, CrSb demonstrates higher phase transition temperature and effective magnetic moment, resulting in a higher spin torque conductivity[24,38]. Hence, CrSb stands out as an ideal material for spintronic application. Previous investigations into the electronic structure of CrSb have been conducted on thin films using soft X-ray ARPES. The limited energy and momentum resolution of the data precluded a clear observation of the anticipated anisotropic and giant SS of the bands[39]. Here, we employ high-resolution vacuum ultraviolet (VUV) and soft X-ray ARPES to probe the SS in CrSb. Our experimental results reveal the evolution of the anisotropic SS in the three-dimensional (3D) momentum space, providing spectroscopic evidence for CrSb as a spin-split antiferromagnet, and showcasing its potential for applications in the evolving landscape of antiferromagnetic spintronics.

The crystal structure of CrSb is shown in Figure 1a). It crystallizes in the NiAs-type crystal structure with the space group *P*6$_3$/*mmc* (#194). The Cr atom is surrounded by six Sb atoms, forming an octahedron with 2-fold rotational symmetry. Below the Néel temperature, it exhibits an out-of-plane *A*-type AFM ground state where the spins align ferromagnetically in-plane and antiferromagnetically between adjacent layers. Figure 1b) displays the spin-polarized electronic density distribution map of CrSb. The electronic density between two Cr atoms cannot be linked through inversion or translation. Instead, the sublattices are linked by the spin group element [$C_2 \parallel M_z$] or [$C_2 \parallel C_{6z}t$] for adjacent Cr atoms along the out-of-plane or in-plane direction (R$_i$ and R$_j$ in [R$_i \parallel$ R$_j$] denote operations in the spin and lattice space, respectively). This symmetry lifts the Kramers degeneracy, resulting in the nonrelativistic SS.

The 3D and surface-projected Brillouin zone (BZ) of CrSb, together with the sign of the *c*-direction spin polarization ($S_z$) in the 3D BZ, are shown in Figure 1c) (the blue/red triangular prisms represent the negative/positive spin components). [$C_2 \parallel M_z$] gives rise to one horizontal mirror plane, enforcing that the bands are degenerated at the *ΓMK* and the *ALH* planes; [$C_2 \parallel C_{6z}t$] gives rise to three vertical mirror planes, indicating that bands are degenerated within the *ΓKHA* plane. The sign of $S_z$ is antisymmetric about the mirror planes, forming an alternative pattern in the 3D BZ.

The DFT-calculated *E-k* dispersion with SOC along the high-symmetry momentum directions *M-Γ-K* and *Γ-A*, together with that along the non-high-symmetry momentum direction *B-O-C* and *E-D*, are shown in Figure 1d). The high-symmetry momentum directions manifest no out-of-plane SS, while significant SS is seen to exist along non-high-symmetry momentum directions *B-O* and *E-D*. The magnitude of SS can reach up to 1.1 eV. Though antimony is expected to have a pronounced SOC effect on the electronic structure, SOC does not have a significant impact on the magnitude of SS (Figure S4, Supporting Information). Therefore, this substantial SS is induced overwhelmingly by the antiferromagnetic order.

The single crystals of CrSb were grown by the chemical vapor transport (CVT) method. The core-level photoemission spectrum reveals the occupied 4*d* orbitals of the Sb atoms and the 3*p* orbital of the Cr atoms [Figure. 1e)]. In Figure 1f), the X-ray diffraction (XRD) results reveal the (00*l*) peaks of CrSb, consistent with the Laue X-ray diffraction pattern. These results are indicative of the high quality of the crystals.

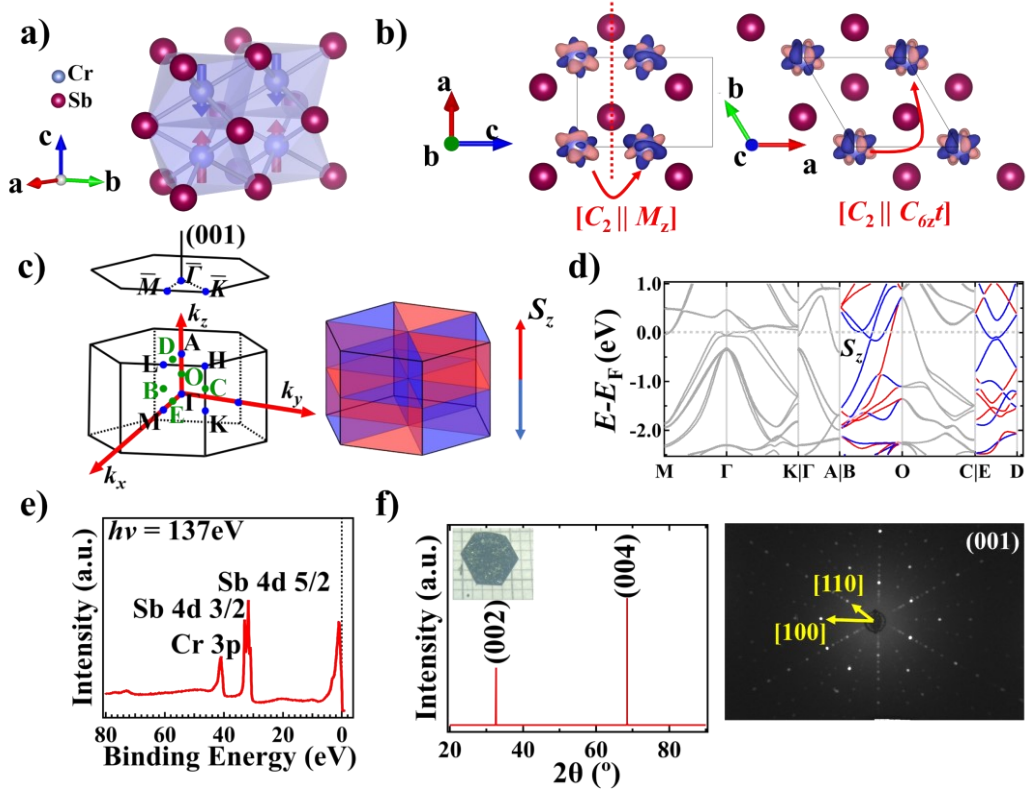

**Figure 1.** High-Néel-temperature metallic spin-split antiferromagnet CrSb. a) Crystal structure of CrSb along with magnetic moments. Red/blue arrows indicate the spin-up and -down orientations of the Cr atom. b) Spin-polarized electronic density distribution of CrSb. Black lines indicate one unit cell. Transposing symmetries $[C_2 \parallel M_z]$ and $[C_2 \parallel C_{6z}t]$ are shown. c) Left: the 3D and surface-projected Brillouin zone (BZ) of CrSb. Blue dots represent high symmetry points; green dots represent the midpoint between two high symmetry points. Right: Schematic for the sign of $S_z$ polarization in the 3D BZ. d) DFT-calculated spin-resolved bands with SOC. There is no spin polarization within the $\Gamma MK$ and $\Gamma KHA$ plane, whereas a high level of spin polarization is seen in the OBC and $\Gamma MLA$ plane. e) Core-level photoemission spectra taken on an *in-situ* cleaved sample at $h\nu = 137$ eV. f) Single crystal XRD results and representative Laue X-ray diffraction pattern of the (001) surface of CrSb. Yellow arrows in the Laue pattern show the orientation indexes of [100] and [110]. Inset: a hexagonally-shaped CrSb single crystal with exposed (001) plane against a millimeter gird.

## 3. Results and Discussion

As mentioned above, the bands are degenerated at $k_z = n\pi/c$ ($n = 0, 1, 2, 3, \ldots$) and split at positions where $k_z \neq n\pi/c$ and ($k_x$, $k_y$) are not within the $\Gamma KHA$ planes. Therefore, the periodic evolution of SS can be revealed within the $\Gamma MLA$ plane and the $OBC$ plane. To illustrate the anisotropic SS within the $\Gamma MLA$ plane, we carried out systematic $h\nu$-dependent ARPES measurements with in-plane momentum aligned along $\bar{\Gamma}$-$\bar{M}$ (Figure 2). The $k_z$-$k_x$ constant energy contours (CECs) demonstrate clear periodic out-of-plane dispersion [Figure 2b]. Through comparison with computational results, the inner potential $V_0$ was determined to be 12.8 eV. The $E_B$-$k_x$ dispersion of CrSb at different $k_z$'s [designated as cuts 1-5 in Figure 2a)] are shown in Figure 2c). The four rows in Figure 2c) correspond to the raw data, the second derivative analysis along the energy distribution curves (EDCs), the DFT-calculated bulk bands with SOC, and the momentum distribution curves (MDCs) along $\bar{M}$-$\bar{\Gamma}$-$\bar{M}$. Within these ARPES spectra, we identify two bands α and β. Even with a rough examination of the raw data in the first row, we can clearly observe that these bands are degenerated at the bulk $\Gamma$ point ($h\nu$ = 118 eV, $k_z$ = 10 $\pi/c$). As $k_z$ moves away from the $\Gamma$ point, the band splits into the α and β bands. At the center of $\Gamma$-$A$ ($h\nu$ = 106 eV, $k_z$ = 9.5 $\pi/c$), the SS is maximal. As $k_z$ continues to decrease, the SS gradually decreases until the two bands remerge into one at the $A$ point ($h\nu$ = 94 eV, $k_z$ = 9 $\pi/c$). To better visualize the SS, we perform the 2$^{nd}$ derivative analysis along the EDCs and fit peaks of the raw MDCs, as shown in the second and the fourth row, respectively. Both analyses demonstrate good agreement between our experimental results and theoretical calculations [the 3$^{rd}$ row in Figure 2(c)], confirming the observation of giant anisotropic SS along the out-of-plane direction. In cut 3, where the maximum SS along $\bar{\Gamma}$-$\bar{M}$ occurs, the energy scale of it is ~0.8 eV at $k_x$ = ± 0.439 Å$^{-1}$, significantly higher than the largest known giant SOC-induced SS in BiTeI (~0.2 eV)[40] and GeTe (~0.2 eV)[41]. Some regions in the 2$^{nd}$ derivative diagram show additional spectral weight, which may result from the $k_z$ broadening effect that reflects bands from neighboring $k_z$'s (Section V of the Supporting Information).

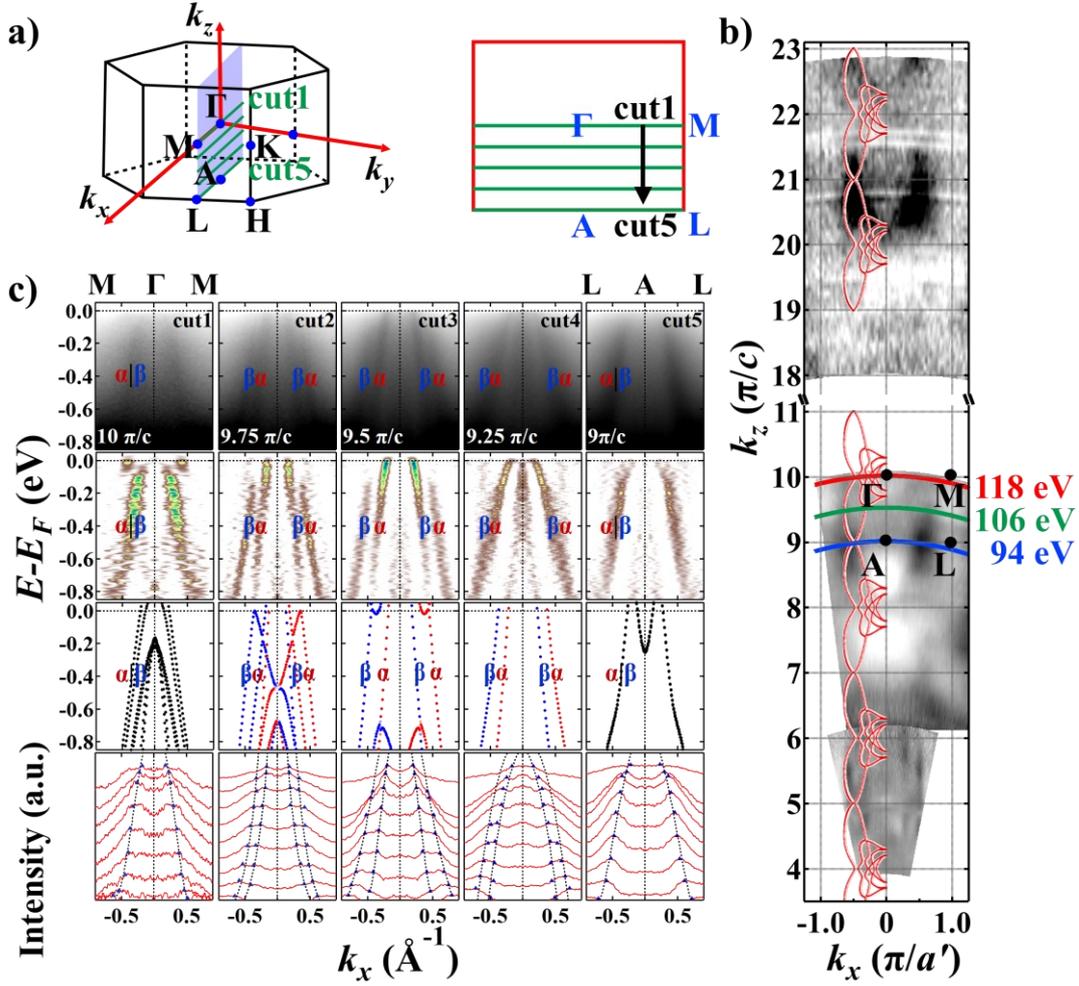

**Figure 2.** Evolution of the band splitting at different out-of-plane momentum positions. a) 3D BZ of CrSb with a 2D cross-section showing the $\Gamma MLA$ plane. The green lines represent measurement positions for cuts 1-5. b) ARPES $k_z$-$k_x$ CECs at binding energies $E_B$ = 0.6 eV, $a' = (\sqrt{3}/2)a$. The inner potential was set to be $V_0$ = 12.8 eV. High symmetry points and $k_z$-$k_x$ curves correspond to three typical photon energies are labeled. The appended red curves represent the DFT calculational results. c) from top to bottom: ARPES band dispersion parallel to $\bar{\Gamma}$-$\bar{M}$ with an equal $k_z$ offset of 0.25 $\pi/c$ between each; corresponding second derivative analysis along the EDCs; DFT-calculated bulk bands (all calculated bands are shifted up by 0.15 eV to account for the charged defects in real crystals), where red and blue correspond to out-of-plane spin-up and -down polarization, respectively; MDCs with peak fitting results. MDCs are drawn with an energy offset of 0.1 eV from $E_B$ = 0 eV to $E_B$ = 0.8 eV. Blue triangles denote the MDC peaks, and the black dashes represent the fitting curves.

**Figure 3.** Evolution of the band splitting at different in-plane directions. a) 3D BZ of CrSb with a 2D cross-section showing the *OBC* plane ($h\nu$ = 106 eV, $k_z$ = -0.5 $\pi/c$). Green lines represent the measurement directions of cuts 1-5 in Panel (c). b) ARPES CECs at $E_B$ = 0, 0.2 , and 0.4 eV with incident $h\nu$ = 106 eV, and DFT calculated results. The $k_x$ direction is aligned along $\bar{\Gamma}$-$\bar{M}$ . c) From top to bottom: ARPES band dispersion along cuts 1-5; corresponding second derivative analysis along the EDCs; DFT-calculated bulk bands along cuts 1-5; MDCs with peak fitting results. MDCs are drawn with an energy offset of 0.1 eV from $E_B$ = 0 eV to $E_B$ = 0.8 eV. Blue triangles denote the MDC peaks, and the black dashes represent the fitting curves.

To illustrate the anisotropic SS within the *OBC* plane, in Figure 3 we carried out ARPES measurements with $h\nu$ = 106 eV [$k_z$ = 9.5 $\pi/c$ ≡ -0.5 $\pi/c$]. The CECs at $E_B$ = 0, 0.2, and 0.4 eV are shown in Figure 3b), where we see that the two triangle-like bands α and β form a hexagram-like pattern resembling the star of David. As the binding energy increases, the star gradually enlarges, exhibiting the characteristics of hole bands. From the calculational results shown in the bottom panel of Figure 3b), the α and β bands possess different spin polarization, revealing that the SS is also anisotropic within the (001) plane. To illustrate the evolution of in-plane SS more clearly, we perform several $E_B$-$k_{//}$ cuts along different in-plane directions [Figure 3c)]. The azimuth angles of cuts 1-5 with respect to $\bar{\Gamma}$-$\bar{M}$ are 0°, 10°, 30°, 50° and 60°, respectively. We observe that when the cut rotates from $\bar{\Gamma}$-$\bar{M}$ (cut 1) to $\bar{\Gamma}$-$\bar{K}$ (cut 3) and to another $\bar{\Gamma}$-$\bar{M}$ (cut 5), the α and β bands move closer, merges into one (along $\bar{\Gamma}$-$\bar{K}$), and split again into the α and β bands. Such periodic variation conforms to the symmetry operation [$C_2 \| C_{6z}t$] of an SS antiferromagnet defined by the spin group shown in Figure 1b).

To resolve the giant SS within the $\Gamma MLA$ plane, we used the self-flux method to grow another set of CrSb single crystals that terminate along the (100) plane. In Figure 4 we carried out systematic $h\nu$-dependent ARPES measurements with in-plane momentum aligned along $\bar{\Gamma}$-$\bar{A}$ to pinpoint the bulk high-symmetry points. The $k_x$-$k_z$ CEC at $E_B$ = 0.6 eV is depicted in Figure 4b). Analogous to observations on the (001) plane, the bands are found to be degenerated at the bulk $\Gamma$ point ($h\nu$ = 100 eV, $k_z$ = 6$\pi/a$'), while split at non-high-symmetry momentum points ($h\nu$ = 82 eV, $k_z$ = 5.5$\pi/a$'), as illustrated in Figure 4c). The magnitude of SS along $\bar{E}$-$\bar{D}$ is measured to be also around 0.8 eV [green dashed line in Figure 4c)]. Although DFT calculations predict an even larger SS of about 1.1 eV, the "spin-up" band (red) close to $E_F$ is not seen in our ARPES data.

In an ideal scenario, similar to MnTe$_2$[30], utilizing spin-resolved ARPES to directly visualize the spin polarization texture would offer the most intuitive demonstration of the unique spin properties in CrSb. However, possible formation of magnetic domains, each much smaller than the size of the beam spot, presents challenges in obtaining reliable spin-resolved spectroscopic data. Nonetheless, it is still possible to distinguish the spin structure of SS antiferromagnets from other signatures of the bands using spin-integrated ARPES. Firstly, bands do not split at high symmetry points, which rules out the possibility of Zeeman SS. Subsequently, since CrSb crystals are centrosymmetric, the bulk Dresselhaus and Weyl SS are prohibited[42–50]. The surface Rashba effect, whose SS has the same magnitude along the out-of-plane direction, contradicts our experimental findings shown in

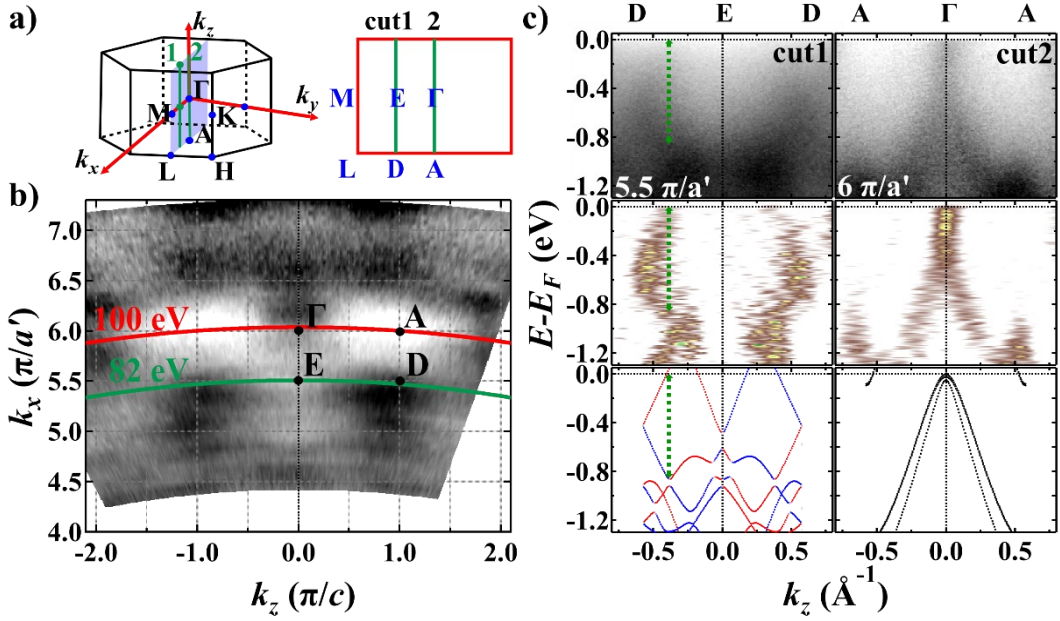

**Figure 4.** Band splitting along non-high-symmetry momentum direction *E-D*. a) 3D BZ of CrSb with a 2D cross-section showing the *ΓMLA* plane. Green lines represent the measurement directions of cuts 1-2 in Panel (c). b) ARPES CEC at $E_B$ = 0.6 eV. The inner potential was set as $V_0$ = 11.8 eV. High symmetry points and $k_x$-$k_z$ curves correspond to two typical photon energies are labeled. c) from top to bottom: ARPES band dispersion along cuts 1-2 with incident $h\nu$ = 82 eV ($k_z$ = 5.5 π/$a'$) and 100 eV ($k_z$ = 6 π/$a'$), respectively; corresponding second derivative analysis along the EDCs; DFT-calculated bulk bands (all calculated bands are shifted up by 0.31 eV to account for the charged defects in real crystals), The green dashed line mark the biggest-seen energy separation of the split bands (~0.8 eV).

Figure 2 that SS becomes zero at the high symmetry planes. According to our calculational results, the sign of spin polarization reverses with a reversal of $k_z$, and the size of the SS varies along both the in-plane and out-of-plane directions. This signifies a spin texture different from the classical Rashba and Dresselhaus SS which are governed by linear terms of $k$[51,52]. Our ARPES data prove the latter of the two characteristics, that SS exists at non-high-symmetry $k_z$'s but has nodes at the bulk *ΓMK* plane. Thus, we conclude that our data reveals a new type of SS induced by unconventional antiferromagnetism.

## 4. Summary and Outlook

In summary, we have successfully conducted a systematic study of the 3D anisotropic SS band structure of CrSb utilizing synchrotron-based VUV and soft X-ray ARPES. The SS of CrSb shows clear $k_z$ dependence and reveals significant anisotropy along in-plane directions. The ARPES-measured energy separation of the split bands is no less than 0.8 eV. Such electronic structure matches our theoretical results, yielding direct, high-resolution spectroscopic evidence for the existence of significant AFM-induced SS in unconventional antiferromagnet CrSb.

Compared to the much-studied semiconducting candidate MnTe, CrSb is a metal with a Néel temperature much higher than the room temperature (~703 K), making it more favorable for devices that are functional at room temperature. Moreover, CrSb exhibits a greater SS strength than what is reported in MnTe (~0.37 eV)[36], indicating that it is more easily controllable by electric fields. Recently, the 100% field-free switching of the Néel vector was observed in CrSb[38]. This demonstrates the great potential of using CrSb to design magnetic-field-free high-density memories. Consequently, our work establishes a firm basis for further studies on potential spintronic applications based on the unique properties of CrSb.

## 5. Experimental Section

*Sample growth and XRD characterization:* Single crystals of CrSb with (001) cleavage plane were grown using the chemical vapor transport method. Starting elements (Cr powder from Aladdin, 99.5% purity; Te powder from Aladdin, 99.5% purity; $I_2$ from Aladdin, 99.9% purity) were grounded and mixed thoroughly in the agate mortar with a molar ratio of Cr: Sb: $I_2$= 1 :1: 0.1. The mixture was then sealed into a silica tube under vacuum. The sealed ampoule was heated in a two-zone furnace to a low-temperature $T_L$= 750 °C and a high-temperature $T_H$ = 850 °C in 12 h and maintained at this condition for two weeks. Millimeter-sized hexagonal-shaped CrSb single crystals were then obtained. Single crystals of CrSb with (100) cleavage plane were grown using the self-flux method. Starting elements (Cr plates from Aladdin, 99.5% purity; Te ingots from Aladdin, 99.5% purity) were packed into an alumina crucible with a molar ratio of Cr: Sb = 3 : 7. The crucible was then sealed in a quartz ampoule under vacuum. The sealed ampoule was heated for 6 h up to 1000 °C, held for 20 h, then slowly cooled down to 750 °C over 100 h, at which temperature the excess Sb-flux was removed by centrifugation. Millimeter-sized needle-like CrSb single crystals were then obtained [Figure S1a), Supporting Information]. The CrSb samples were characterized by XRD at room temperature using a Rigaku SmartLab diffractometer with Cu Kα radiation. The diffraction pattern in Figure 1f) confirms that the cleaving planes of the first type of crystals are parallel to the (001) crystallographic orientation. The XRD result of CrSb samples with (100) cleavage plane is shown in Figure S1.

*Energy Dispersive X-Ray Spectroscopy Measurement:* EDX experiment was operated on Nova NanoSem450 at an accelerating voltage of 20 kV and a current of 1 nA.

*Laue X-ray Diffraction Measurement:* Laue X-ray Diffraction measurement was performed on a Laue crystal orientation system (LCS2020W) designed by the Shanghai Institute of Ceramics, Chinese Academy of Sciences. During the measurement, the sample was positioned at a distance of 5 cm from the X-ray source. The exposure time was set to 120 seconds to ensure an adequate signal-to-noise ratio.

*Magnetic Characterization:* Magnetization measurements of the CrSb single crystals were carried out with a superconducting quantum interference device (SQUID)-vibrating sample magnetometer (VSM) system (MPMS3, Quantum design). This system is capable of cooling samples down to 1.8 K and can generate a variable magnetic field up to ±7 T along both in-plane and out-of-plane directions. FC and ZFC curves were measured by increasing the temperature from 2 to 400 K with both in-plane and out-of-plane magnetic fields of 500 Oe.

*ARPES Measurements:* The $k_z$ dispersion data and the spin-integrated electronic structure at different photon energies were performed at BL03U[53] and BL09U of the Shanghai Synchrotron

Radiation Facility (SSRF), BL09A of the Hiroshima Synchrotron Radiation Center (HiSOR) and BL25SU of the SPring-8 synchrotron facility. Data at BL03U of the SSRF are measured with a Scienta Omicron DA30 electron analyzer and *p*-polarized light with photon energies between 40 and 120 eV. Data at BL09U of the SSRF are measured with a Scienta Omicron DA30 electron analyzer and *p*-polarized light with photon energies between 94 and 144 eV. Data at BL09A of the HiSOR are measured with a SPECS PHOIBOS 150 electron analyzer and *p*-polarized light with photon energies between 11 and 40 eV. Data at BL25SU of SPring-8 are measured with a Scienta Omicron DA30 electron analyzer and $C^+$-polarized light with photon energies between 400 and 650 eV. The measurement temperature is set to around 30 K at SSRF and HiSOR and around 77 K at SPring-8. The samples for all ARPES measurements are cleaved *in-situ* and measured in a vacuum better than $2 \times 10^{-10}$ Torr.

*First-Principles Calculations*: The electronic structure calculations were carried out using the DFT method encoded in the Vienna Ab-initio Simulation Package (VASP)[54,55] based on the projector augmented wave (PAW) method[56]. The Perdew-Burke-Ernzerhof (PBE) approximation is used for the exchange-correlation function[57]. The plane-wave cutoff energy was set to 520 eV. GGA+U correction was applied to the Cr 3*d* orbitals, and U was set to be 1.0 eV. The *k*-point sampling is 10 × 10 × 6 with the $\Gamma$ scheme for the bulk structure. To study the CECs of CrSb along (001), We determined maximally localized Wannier functions using a reduced basis set formed by the *d* orbitals of Cr, *s*, and *p* orbitals of Sb atoms in the Wannier90 software[58]. We used the WannierTools package to simulate the theoretical CECs[59]. The experimental values used for cell parameters are *a* = *b* = 4.18 Å, and *c* = 5.46 Å. Atomic positions are fully relaxed until the force on each atom is smaller than $1\times10^{-3}$ eV/Å, and the total energy convergence criterion is set to be $1\times10^{-7}$ eV.

## Acknowledgments

We thank Yang Liu and Yuntian Liu for helpful discussions, Taichi Okuda, Kenta Kuroda, Yudai Miyai, and Cheng Zhang for the help in VUV-ARPES measurements in HiSOR, and Dawei Shen, Wenchuan. Jing for the help in VUV-ARPES measurements in BL03U of the SSRF. Work at SUSTech was supported by the National Key R&D Program of China (nos. 2022YFA1403700 and 2020YFA0308900), the National Natural Science Foundation of China (NSFC) (nos. 12204221 and 12274194), the Key-Area Research and Development Program of Guangdong Province (2019B010931001), the Guangdong Provincial Key Laboratory for Computational Science and Material Design (no. 2019B030301001), the Guangdong Innovative and Entrepreneurial Research Team Program (no. 2016ZT06D348), the Guangdong Natural Science Foundation (No. 2022A1515012283), and the Shenzhen Science and Technology Program (grant no. RCJC20221008092722009). The DFT calculations were performed at the Center for Computational Science and Engineering at SUSTech. The ARPES experiments were performed at BL03U and BL09U of Shanghai Synchrotron Radiation Facility under the approval of Proposals 2023-SSRF-PT-502004 and 2022-SSRF-PT-020799-1, at BL09A in HiSOR under the approval of Proposal No. 23BG002, and at BL25SU in SPring-8 under the approval of Proposal No. 2023A1187.

## Keywords